\documentclass[fleqn,twoside]{article}
\usepackage{espcrc2}
\usepackage{epsf}
\usepackage[figuresright]{rotating}
\def\be{\begin{equation}}
\def\ee{\end{equation}}

\newcommand{\AmS}{{\protect\the\textfont2
  A\kern-.1667em\lower.5ex\hbox{M}\kern-.125emS}}
\title{Large N QCD -- Continuum reduction}

\author{Rajamani Narayanan
\address[FIU]{Department of Physics, University Park, Miami, FL 33199}}
       
\begin{document}

\begin{abstract}

Numerical evidence combined with Eguchi-Kawai reduction indicate that
there are no finite volumes effects in the large N limit of QCD as
long as the linear extent of the four-torus is bigger than a critical
size. This is referred to as continuum reduction.
Since fermions in the fundamental representation are naturally
quenched in the large N limit, as long as we only have a finite number
of flavors, continuum reduction provides us with the exciting
possibility to numerically solve large N QCD using chiral fermions
and present day computers.
\vspace{1pc}
\end{abstract}

\maketitle

\section{Introduction}

Overlap fermions~\cite{overlap} enables one to compute quantities in the chiral
limit of lattice QCD. But one has to work with $\epsilon(\gamma_5 D_w(-m))$
where $D_w(-m)$ is the Wilson-Dirac operator in the supercritical region
and $\epsilon$ is the sign function of the operator. This makes the
algorithms on the lattice computationally intensive and the situation
is further complicated by the presence of very small eigenvalues of
$\gamma_5 D_w(-m)$~\cite{wilson}. 
In spite of these difficulties several results have
been obtained in the quenched approximation~\cite{giusti} and 
dynamical simulations using overlap fermions have been explored~\cite{bhen}. 

Large N QCD~\cite{hooft1} provides an interesting alternative to dynamical
simulations of SU(3) QCD since fermions in the fundamental representation
are naturally quenched in the $N\rightarrow\infty$ limit of large N
QCD as long as the number of flavors are kept finite. 
It is useful to use the ``double line'' notation for the
gauge propagator in large N QCD to denote that there are two color
indices associated with the adjoint representation of the gauge field.
In this notation, fermions in the fundamental representation have
only one line in their propagator. Only planar diagrams survive in the
limit of $N\rightarrow\infty$ and the vacuum diagrams arising out
of the fermionic part are suppressed by a factor of $\frac{1}{N}$
and fermions are naturally quenched~\cite{hooft1}. 
Large N QCD has all the qualitative features of $N=3$ QCD and the
spectrum of  mesons has been calculated in two dimensions~\cite{twod}.
One also expects a non-trivial spectrum of mesons in the large N limit
of four dimensional QCD but this model has not yet been exactly solved.

\section{Continuum reduction}

Eguchi and Kawai~\cite{ek} proved that lattice QCD
in the limit of $N\rightarrow\infty$ can be reduced to QCD on a single
site provided the global $U^d(1)$ symmetries that multiplies
the link matrices by $U(1)$ phases are not broken. Unfortunately,
this symmetry
is broken for $d>2$~\cite{bhn} and one cannot reproduce QCD in
the large N limit by working on a $1^d$ lattice.
But, the proof of Eguchi and
Kawai also holds for an $L^d$ lattice where $L >1$. That is to say,
the infinite space-time lattice can be reduced to a finite $L^d$
lattice in the limit of $N\rightarrow\infty$ provided the
global $U^d(1)$ symmetries that multiplies
Polyakov loops by $U(1)$ phases are not broken on the $L^d$ lattice.

The only parameter in the lattice theory is the 't Hooft gauge coupling
$b=\frac{1}{g^2N}$. This coupling has the dimensions of length in
three dimensions and it is dimensionless in four dimensions. One can
explicitly compute the location of the transition point $b_c(L)$
such that the $U^d(1)$ symmetries remain unbroken for $b<b_c(L)$.
If $b_c(L)$
scales properly with $L$ then one can define a critical size
$l_c$ in the continuum such that the $U^d(1)$ symmetries remain
unbroken for $l>l_c$. The argument of Eguchi and Kawai will hold for
all $L$ as long as we keep $b<b_c(L)$.
There will be no dependence on the box size
as long as the box size is bigger than the critical size. This is
referred to as continuum reduction~\cite{threed}.
Physical
results can be extracted by working on an $L^d$ lattice and keeping
$b$ just below $b_c(L)$. Computations need to be done on two or three
different values of N to study the infinite N limit. 
Results for two or three different $L$
values can be used to study the effect of finite lattice spacing.

\section {The bulk transition}

One lattice artifact has to be taken into account when setting
the parameters $\{L,b<b_c(L)\}$ for physical computation. 
There is a bulk transition on the
lattice in the large N limit that is associated with the spectrum
of the single plaquette ($1\times 1$ Wilson loop). There
is no phase transition for finite N but there is a cross-over
that gets stronger with increasing N. The eigenvalues
$e^{i\theta}$ of $U_p$, the parallel transporter along the
plaquette, can be anywhere in the range $-\pi \le \theta < \pi$. 
But the field strength is small close to the continuum limit and
$\theta$ is expected to be in a small range close to zero.
In fact, there is a value of coupling $b_B$ such that the distribution
of $\theta$ develops a gap around $\pm\pi$ for $b>b_B$. This value
of coupling does not depend on $L$ and therefore the transition
is purely a lattice artifact. But one has to set $b> b_B$ to obtain
the proper continuum limit. Only for some $L>L_B$ does $b_c(L)$ become
bigger than $b_B$. For $L\le L_B$ all one has is the bulk transition
and the $U^d(1)$ symmetries also break at the same coupling. This is
the reason why one cannot obtain continuum QCD in the large N limit
using the single site Eguchi-Kawai model and there is a smallest lattice
size that one has to use in $\{L > L_B ,b_B < b < b_c(L)\}$.

\section {Numerical results}

In order to study QCD in the large N limit, one has to first obtain
the appropriate set of parameters $\{L > L_B ,b_B < b < b_c(L)\}$.
The $U^2(1)$ symmetries are not broken in $d=2$ and $b_c(L)=\infty$
for all $L\ge 1$. The bulk transition occurs at $b_B=0.5$~\cite{gw}
and Eguchi-Kawai reduction holds on a $1^2$ lattice. 

Numerical study involves the generation of thermalized gauge field
configurations and this is obtained using a combination of heat-bath
and over-relaxation updates. The heat-bath updates are performed on
all SU(2) subgroups and the over-relaxation updates are performed
on the full SU(N) group~\cite{knn}.
A
preliminary investigation in three dimensions show that $b_B=0.4$ and
$L_B=2$~\cite{threed}. 
Therefore, one can obtain results in three dimensional large N
QCD by working on as small a lattice as $3^3$. The breaking of the
$U^3(1)$ symmetries occurs as follows. There is a $b_c(L)$ where one
of the three $U^3(1)$ are broken and the other two remain unbroken.
The eigenvalue distribution of the Polyakov lines in the unbroken
direction are flat (random) but there is a peak in the distribution of
the broken line. As one goes above $b_c(L)$ more lines gets ordered
and there are two more transition points at any finite $L$. More
investigation has to performed to study the behavior of the higher
transitions as a function of $L$, but the lowest transition corresponding
to one ordered Polyakov line scales properly with $L$. It is quite
likely that all these transitions fuse as one goes to the continuum limit.
An analysis of
this transition shows that $0.6 < b_c(3) < 0.7$, $0.8 < b_c(4) < 0.9$,
$1.0< b_c(5) < 1.2$ and $1.2 < b_c(6) < 1.35$. When combined, these
results imply 
\be
4.2 < \frac{L}{b_c(L)} < 5 \ \ \ {\rm for} \ d=3.
\ee

\begin{figure}
\begin{center}
\includegraphics[height=7cm,width=7cm]{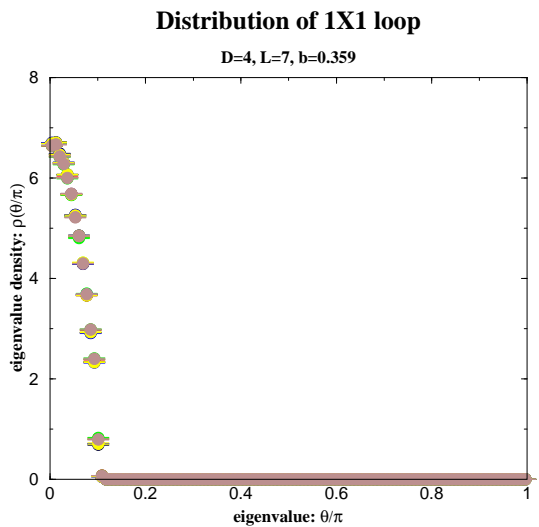} 
\end{center}
\caption{Eigenvalue distribution of the $1\times 1$ Wilson loop
at $b=0.359$ shows a gap.}
\label{wil1a}
\end{figure}

\begin{figure}
\begin{center}
\includegraphics[height=7cm,width=7cm]{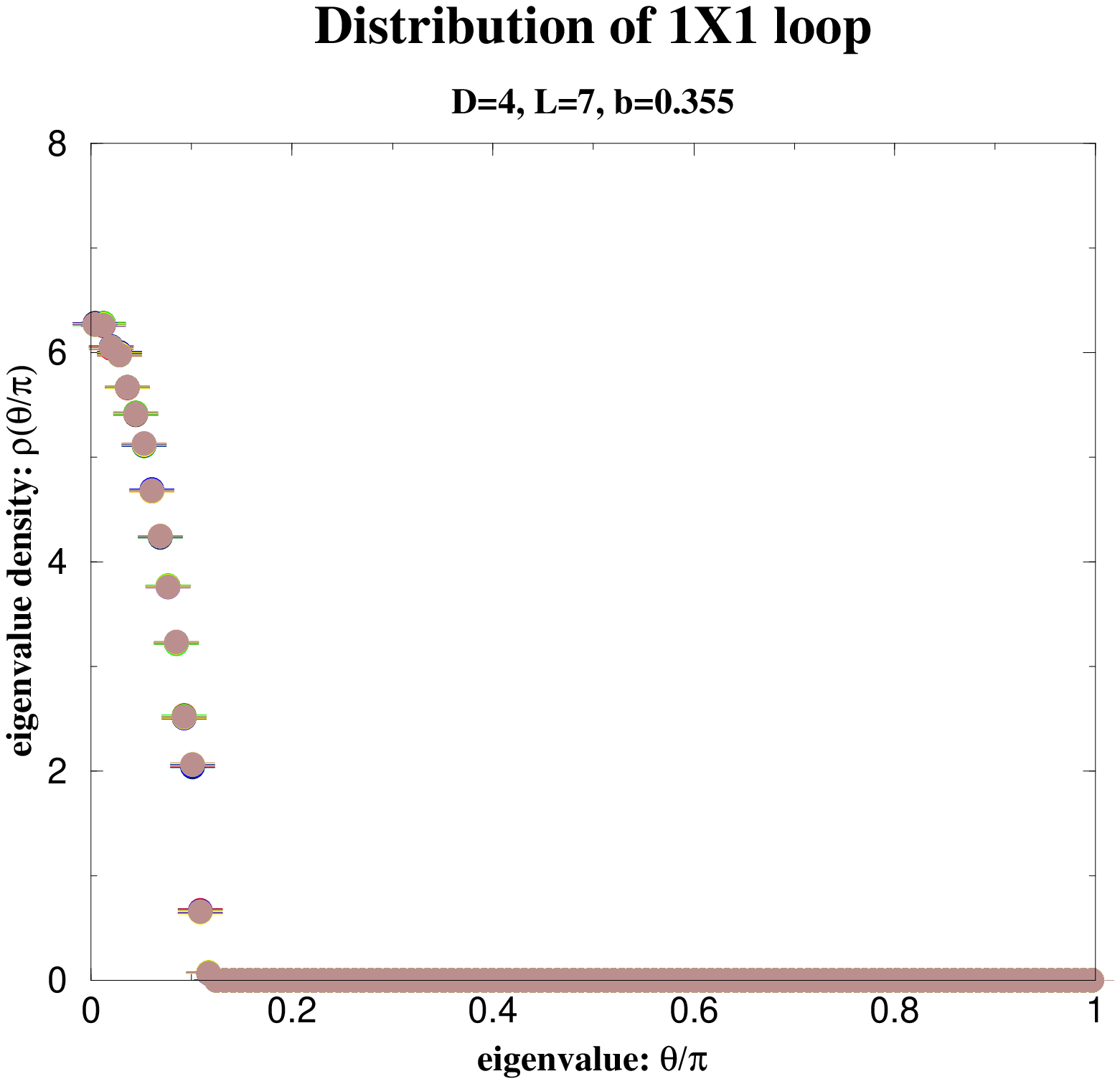}
\end{center}
\caption{Eigenvalue distribution of the $1\times 1$ Wilson loop
at $b=0.355$ shows a gap.}
\label{wil1b}
\end{figure}

\begin{figure}
\begin{center}
\includegraphics[height=7cm,width=7cm]{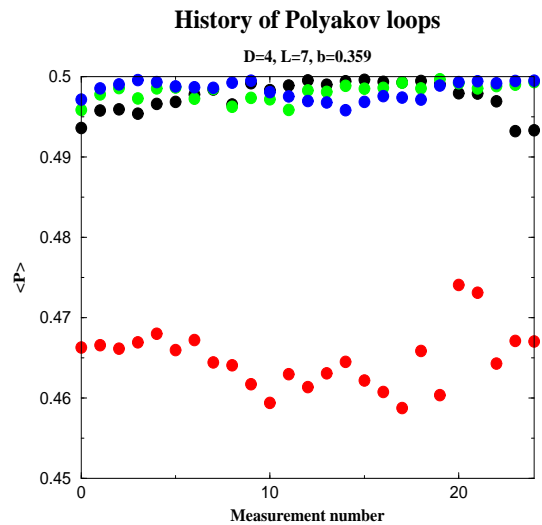}  
\end{center}
\caption{Time history of $p(\tilde P_\mu)$ shows that the
$U(1)$ symmetry is broken in one of the four directions.}
\label{timea}
\end{figure}

\begin{figure}
\begin{center}
\includegraphics[height=7cm,width=7cm]{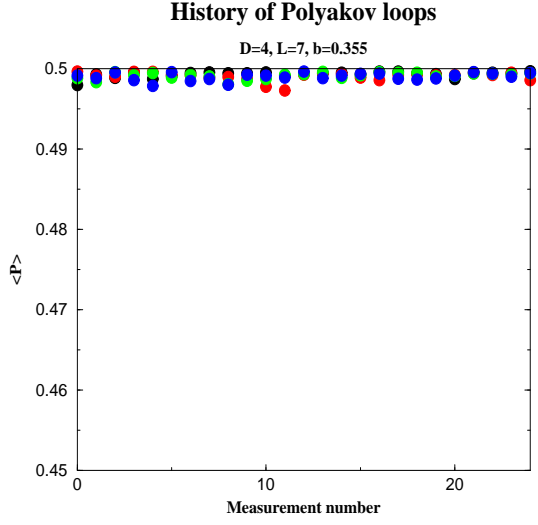}  
\end{center}
\caption{Time history of $p(\tilde P_\mu)$ shows that the
$U(1)$ symmetry is preserved in all four directions.}
\label{timeb}
\end{figure}

\begin{figure}
\begin{center}
\includegraphics[height=7cm,width=7cm]{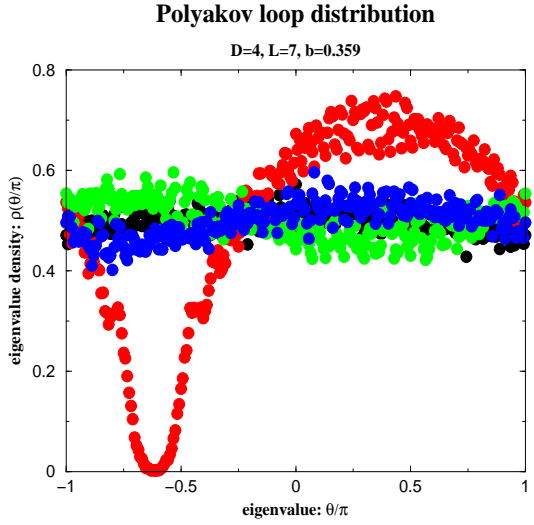}
\end{center}
\caption{Eigenvalue distribution of $\tilde P_\mu$ 
at $b=0.359$ shows that the
$U(1)$ symmetry is broken in one of four directions.}
\label{loopa}
\end{figure}

\begin{figure}
\begin{center}
\includegraphics[height=7cm,width=7cm]{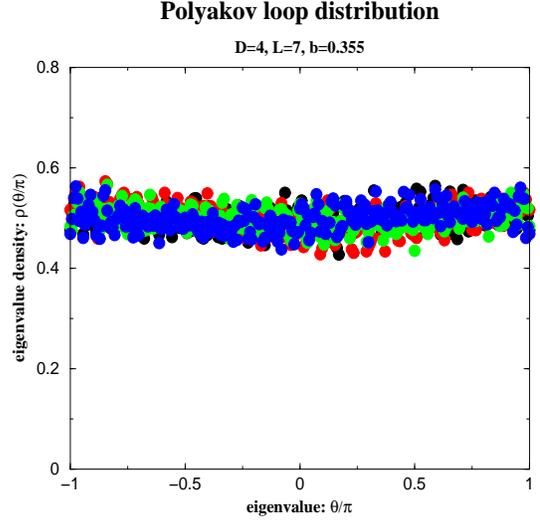}
\end{center}
\caption{Eigenvalue distribution of $\tilde P_\mu$ 
at $b=0.355$ shows that the
$U(1)$ symmetry is preserved in all four directions.}
\label{loopb}
\end{figure}

\begin{figure}
\begin{center}
\includegraphics[height=7cm,width=7cm]{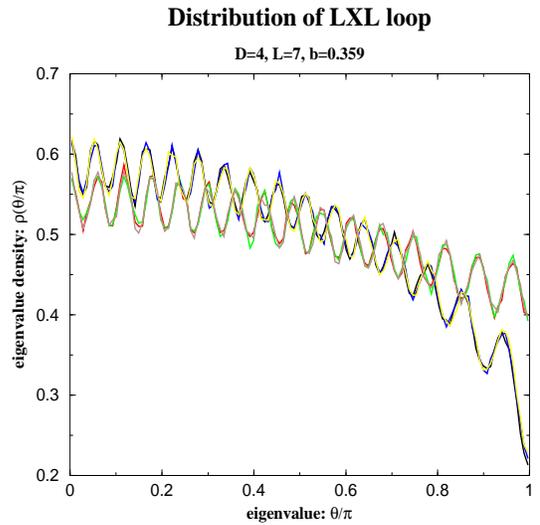} 
\end{center}
\vskip -1cm
\caption{Eigenvalue distribution of the six different orientations
of the $L\times L$ wilson loop
at $b=0.359$ showing that
the $U(1)$ symmetry is broken in one of four directions.}
\label{wil7a}
\end{figure}

\begin{figure}
\begin{center}
\includegraphics[height=7cm,width=7cm]{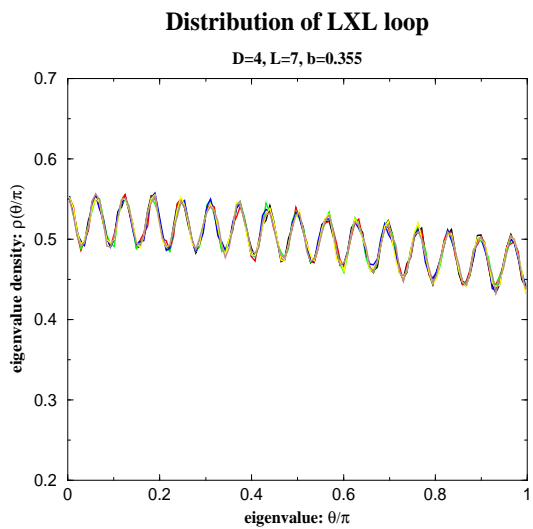}
\end{center}
\vskip -1cm
\caption{Eigenvalue distribution of the six different orientations
of the $L\times L$ wilson loop
at $b=0.355$
indicating that
the $U(1)$ symmetry is preserved in all four directions.}
\label{wil7b}
\end{figure}

\begin{figure}
\begin{center}
\includegraphics[height=7cm,width=7cm]{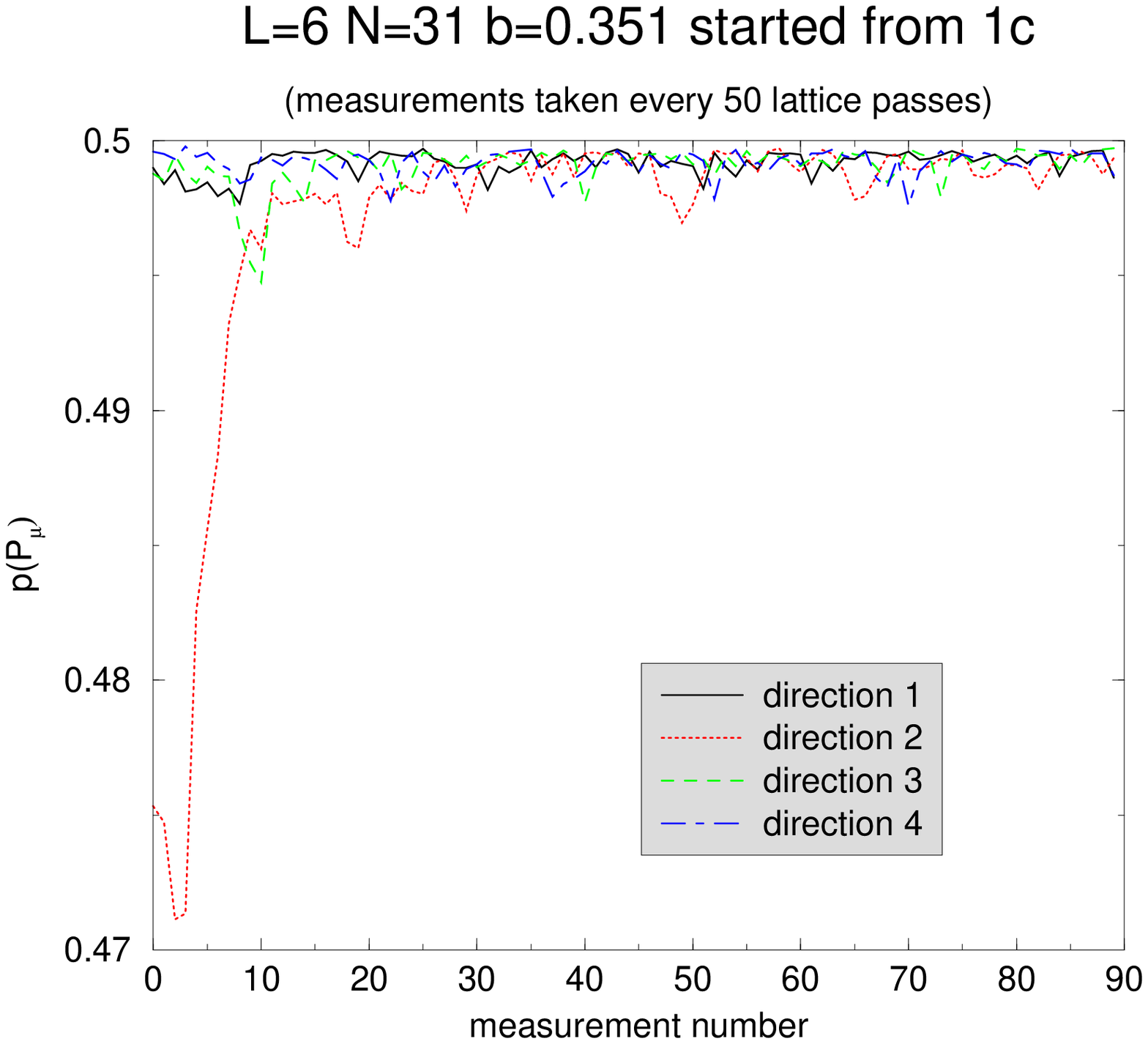}
\end{center}
\caption{ Time history of $p(\tilde P_\mu )$ for each direction
showing the evolution from $1c$ to $0c$.
}
\label{1cto0c}
\end{figure}

\begin{figure}
\begin{center}
\includegraphics[height=7cm,width=7cm]{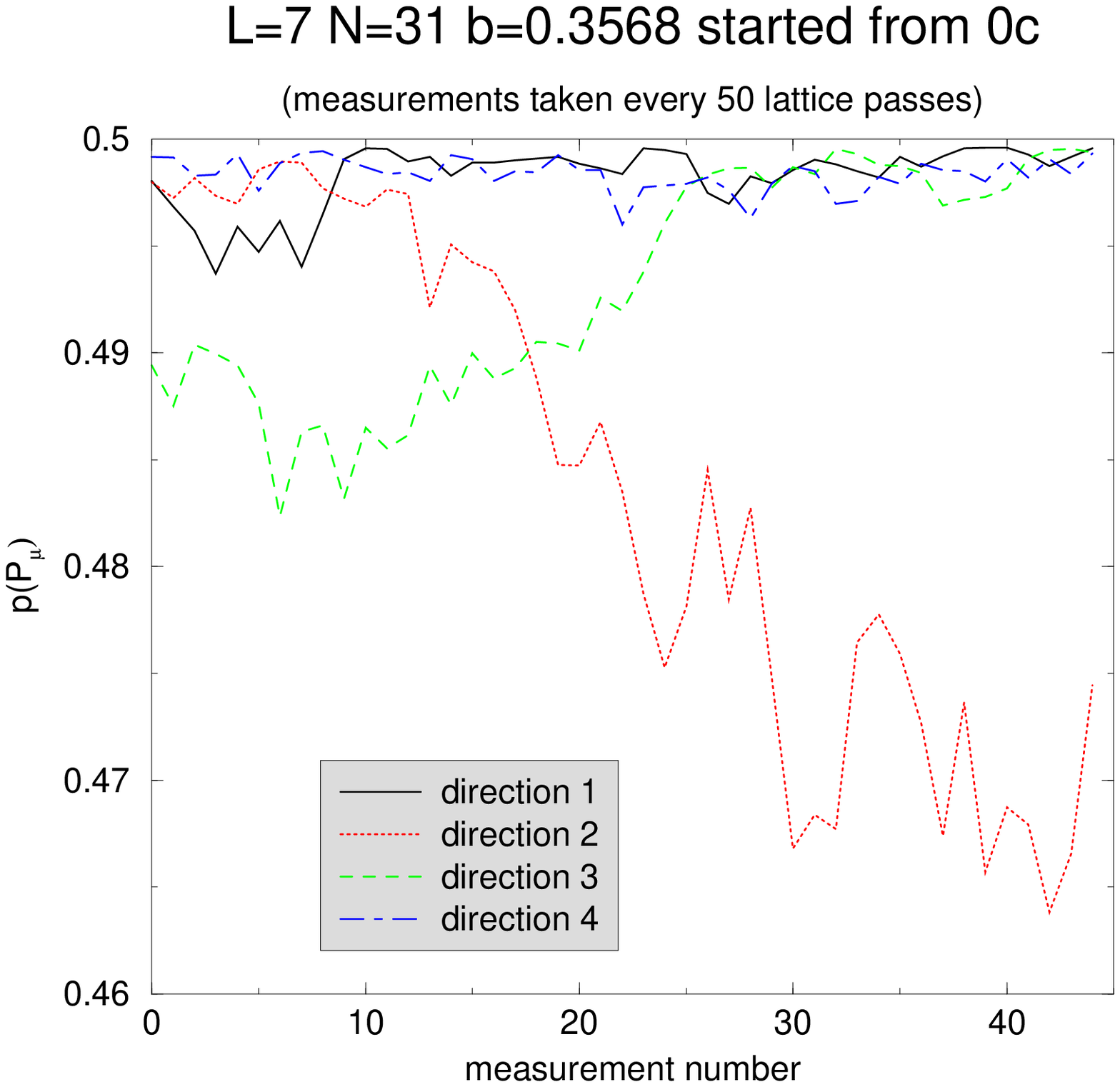}
\end{center}
\caption{ Time history of $p(\tilde P_\mu )$ for each direction
showing the evolution from $0c$ to $1c$.
}
\label{0cto1c}
\end{figure}

A study of the twisted Eguchi-Kawai model
in four dimensions shows that $b_B=0.36$ but that this transition
is strongly first order~\cite{camp}. 
Numerical investigation of the $U^4(1)$ breaking transition shows
that only for $L>7$ does $b_c(L)$ go above $0.36$~\cite{knn}.
Like in three dimensions,
there are many transitions associated with $U^4(1)$ breaking on
the lattice.
Let $0c$, $1c$, $2c$, $3c$ and $4c$ denote phases with $0$,$1$,$2$,$3$
and $4$ ordered Polyakov loops with $b$ above $b_B$.
Let $0h$ denote the phase below $b_B$.
One can start with a $1c$ phase and bring it down to a $0c$ phase
and still be below $b=0.36$. This is a metastable state.
This metastable transition from $1c\rightarrow 0c$ is shown in
Figs.~\ref{wil1a}-\ref{wil7b}. The system is in a $1c$ state at $b=0.359$
on a $7^4$ lattice with $N=37$ and it goes down to a $0c$ state
at $b=0.355$. We know that both $b=0.359$ and $b=0.355$ are
on the weak coupling side of
the bulk transition since the eigenvalue
distributions of the $1\times 1$
Wilson loop shown in Fig.~\ref{wil1a}
and Fig.~\ref{wil1b} exhibit strong gaps at both values of $b$.
The breaking of the $U^4(1)$ symmetry can be probed by looking
at Polyakov loop observables.
The Polyakov loops (as well as various Wilson loops) 
were built out of ``smeared'' $\tilde U_\mu(x)$ matrices,
rather than the original link matrices $U_\mu (x)$~\cite{knn}
since it suppresses ultraviolet fluctuations.
Let $\tilde P_\mu$ be the smeared Polyakov loop in the $\mu$
direction and let its eigenvalues be $e^{i\theta_i}$; $i=1,\cdots,n$.
The quantity~\cite{bhn}
\begin{equation}
p(\tilde P_\mu) = \frac{1}{N^2} \langle \sum_{i,j=1}^N \sin^2 \frac{1}{2} 
(\theta_i - 
\theta_j )\rangle 
\end{equation}
averaged over the 3-plane perpendicular to $\mu$ will be close to
$0.5$ if the $U(1)$ symmetry in that direction is not broken. Breaking
of $U(1)$ symmetry in the $\mu$ direction will open a gap in the
spectrum of $\tilde P_\mu$ are $p(\tilde P_\mu)$ will drop below $0.5$.
A time history of $p(\tilde P_\mu)$ in all four directions at
$b=0.359$ (Fig.~\ref{timea})
shows that the $U(1)$ symmetry is broken in one of the
four directions but the time history at $b=0.355$ 
(Fig.~\ref{timeb}) shows that all
four $U(1)$ symmetries are preserved. Therefore $b=0.359$ is in the
$1c$ phase and $b=0.355$ is in the $0c$ phase.
Eigenvalue distributions of $\tilde P_\mu$ at $b=0.359$
also exhibit
the breaking of the U(1) symmetry in one of the directions
as seen in Fig.~\ref{loopa}. But all four eigenvalue
distributions of $\tilde P_\mu$ at $b=0.355$ look uniform
as seen in Fig.~\ref{loopb}.
Further evidence for $b=0.359$  in the $1c$ phase
and $b=0.355$ in the $0c$ phase can be seen by looking
at the eigenvalue distribution of the $L\times L$ wilson
loop. This distribution should not have a gap and all
six orientations of the wilson loop should look the
same at $b=0.355$ but three of the six wilson loops
should look different from the three others at $b=0.359$.
This is indeed the case as demonstrated in Fig.~\ref{wil7a}
and Fig.~\ref{wil7b}. The oscillations seen in the distributions
are a finite N effect. Finally, one can see that the system
stabilizes in the correct phase even if it started in the wrong
phase. For example, a system started in the $1c$ phase on
a $6^4$ lattice at $b=0.351$ ends up in the correct $0c$ phase after
a few sweeps of the lattice as seen in Fig.~\ref{1cto0c}.
Similarly, a system started in the $0c$ phase on a $7^4$ lattice
at $b=0.3568$ ends up in the correct $1c$ phase after a few sweeps of
the lattice as seen in Fig.~\ref{0cto1c}.

Such a metastable state can be used to reduce $L$ where one
realizes large N QCD.
We find that one can go as low as $L=5$ and remain in the $0c$ phase.
The $1c\rightarrow 0c$ transitions are located at
$b_c(5)=0.34775\pm 0.00075$,
$b_c(6)=0.35125\pm 0.00075$, and
$b_c(7)=0.3564\pm 0.004$.
For $L>7$, the transition from unbroken $U^4(1)$ to one broken
$U(1)$ occurs in the stable region and $b_c(8)=0.3605\pm 0.0015$,
$b_c(9)=0.364\pm 0.001$, and $b_c(10)=0.367\pm 0.001$.

\begin{figure}
\begin{center}
\includegraphics[height=7cm,width=7cm]{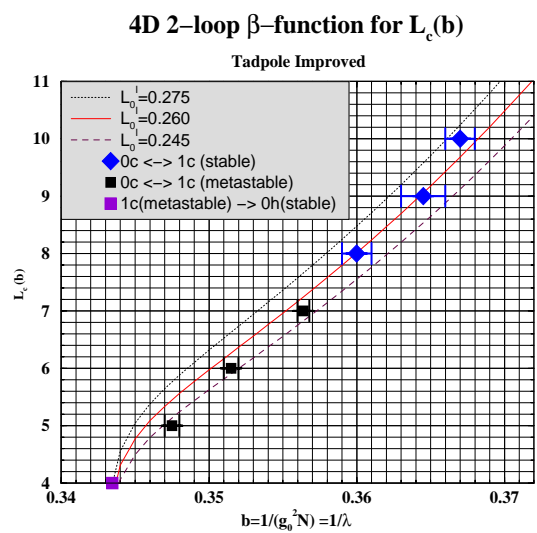}
\end{center}
\caption{Demonstration of scaling in four dimensional large N
QCD}
\label{scale}
\end{figure}

The above values for the critical coupling in the range $L=5$ to $L=10$
can be used to study the scaling of the critical coupling in four
dimensions~\cite{knn}. 
The inverse function $L_c(b)$ should scale according to
\begin{equation}
L_c(b)~\sim ~L_0 \left ( \frac {11}{48\pi^2 b} \right )^{\frac{51}{121}} 
e^{\frac{24\pi^2 b}{11}}
\end{equation}
if it is a truly continuum phenomenon. Since the above result is
a two-loop result, one has to use the
the ``tadpole'' \cite{lepage} replacement
\begin{eqnarray}
b\rightarrow b_I \equiv b e(b)~~~~~~~e(b)=\frac{1}{N} 
\langle Tr U_{\mu,\nu} (x) \rangle \\
L_c(b)~\sim ~L^I_0 \left ( \frac {11}{48\pi^2 b_I }\right )^{\frac{51}{121}} 
e^{\frac{24\pi^2 b_I}{11}}
\end{eqnarray}
to study scaling. Fig.~\ref{scale} shows that the data does
scale properly with
$L^I_0$ between 0.245 and 0.275. 

\section{Conclusions and future work}
The breaking of the $U^d(1)$ symmetries for $d>2$ implies that
Eguchi-Kawai reduction does not work on a $1^d$ lattice. Quenched
Eguchi-Kawai models~\cite{bhn} and twisted Eguchi-Kawai models~\cite{twist}
provide a way out of this problem. The problem of including fermions
in the twisted version of the model has not been completely solved.
It is possible to extract physical quantities in the large N limit
of QCD using the quenched version of the model as shown in two
dimensions~\cite{knn2}. But it is necessary to sample all the
$N^d$ minima in the quenched model and this could prove to be
numerically difficult in four dimensions. An alternative to the
quenched Eguchi-Kawai model is provided by the idea of continuum
reduction. Continuum reduction implies that there are no finite
volume effects in the large N limit of QCD as long as one works
with $L>L_B$ and $b_B < b < b_c(L)$. 

The eigenvalue distribution of the single plaquette exhibits a gap
and the space of gauge field configurations naturally splits into
several disconnected pieces. It is natural to assume that the disconnected
pieces correspond to different gauge field topologies and this can
be explicitly verified by a numerical simulation. Since there are
no finite volume effects, one should see evidence for a chiral condensate
if the box size is bigger than the critical size and the chiral condensate
should occur due to the lowest eigenvalues scaling like $1/N$ as N
gets large. This also can be verified using numerical simulations on
fairly small lattices ($L=8-10$, $N=20-30$).

Since there are no finite volume effects it should be possible to
compute meson propagators for arbitrarily small momenta as long
as the box size is bigger than the critical size. Normally, it
is not possible to work with momenta that are smaller than the
inverse of the box size. This can be avoided in the large N limit
of QCD since a constant $U(1)$ phase multiplying all the link
matrices in a given direction can be interpreted as momentum
carried by a quark line~\cite{bhn,knn,momen}. This interpretation works
as long as the $U^d(1)$ symmetries are not broken since the pure
gauge sector is invariant under a multiplication by a constant phase
but the fermion propagator picks up a momentum equal to this
constant phase. This is indeed the case if the box size is bigger
than the critical size. A direct computation of the meson propagators
in momentum space will lend credibility to the above argument.

\section{Acknowledgments}
All results presented in this contribution to the
LHP2003 workshop are based on collaborative work
with Joe Kiskis and Herbert Neuberger. 
The author acknowledges partial support by the NSF under
grant number PHY-0300065 and also partial support from Jefferson 
Lab. The Thomas Jefferson National Accelerator Facility
(Jefferson Lab) is operated by the Southeastern Universities Research
Association (SURA) under DOE contract DE-AC05-84ER40150.
The author would like to thank the organizers of the LHP2003
workshop for a lively and pleasant atmosphere.


\end{document}